\documentclass[aps,pra,twocolumn,superscriptaddress]{revtex4}
\usepackage{graphicx}
\usepackage{amsfonts}
\usepackage{amsthm}
\usepackage{bm}
\usepackage{color}
\usepackage{epsfig}
\usepackage{amsmath}
\usepackage{amssymb}
\begin{document}

\title{Terahertz Quantum Cryptography}

\author{Carlo Ottaviani}
\email{carlo.ottaviani@york.ac.uk} \affiliation{Computer Science and York Centre for
Quantum Technologies, University of York, York YO10 5GH, UK}
\author{Matthew J. Woolley}
\affiliation{School of Engineering and Information Technology,
University of New South Wales,\\ Australian Defence Force Academy,
Canberra, ACT 2600, Australia.}
\author{Misha
Erementchouk} \affiliation{Department of Electrical Engineering \&
Computer Science, University of Michigan, Ann Arbor, MI, USA.}
\author{John F.
Federici} \affiliation{New Jersey Institute of Technology, 323
King Blvd., Newark, NJ 07102.}
\author{Pinaki
Mazumder} \affiliation{Department of Electrical Engineering \&
Computer Science, University of Michigan, Ann Arbor, MI, USA.}
\author{Stefano Pirandola}\email{stefano.pirandola@york.ac.uk}
\affiliation{Computer Science and York Centre for Quantum
Technologies, University of York, York YO10 5GH, UK}
\affiliation{Research Laboratory of Electronics, Massachusetts
Institute of Technology, Cambridge, Massachusetts 02139, USA}
\author{Christian Weedbrook}
\affiliation{Xanadu, 372 Richmond St. W, Toronto, M5V 2L7,
Canada.}

\date{\today}


\begin{abstract}
A well-known empirical rule for the demand of wireless communication systems
is that of Edholm's law of bandwidth. It states that the demand for bandwidth
in wireless short-range communications doubles every 18 months. With the
growing demand for bandwidth and the decreasing cell size of wireless systems,
terahertz (THz) communication systems are expected to become increasingly
important in modern day applications. With this expectation comes the need for
protecting users' privacy and security in the best way possible.
With that in mind, we show that quantum key distribution can
operate in the THz regime and we derive the relevant secret key
rates against realistic collective attacks. In the extended THz
range (from $0.1$ to $50$~THz), we find that below $1$ THz, the
main detrimental factor is thermal noise, while at higher
frequencies it is atmospheric absorption. Our results show that
high-rate THz quantum cryptography is possible over distances
varying from a few meters using direct reconciliation, to about
$220$m via reverse reconciliation. We also give a specific example
of the physical hardware and architecture that could be used to
realize our THz quantum key distribution scheme.
\end{abstract}
\maketitle

\section{Introduction}

The demand for wireless service data rates has increased exponentially in the
last decade. However, such a trend is fundamentally limited by Shannon's
channel capacity~\cite{Shannon1948}. Beyond this point higher carrier
frequencies must be utilized to provide sufficient data rate
capacity~\cite{Federici2013,Federici2010,Akyildiz2014,Akyildiz2014a,Nagatsuma2013,Kleine-Ostmann2011}%
. There are two competing carrier bands for high data rate ($>100$~Gb/s)
wireless links. The most well-known option is free-space optical
communications. The second option is to increase the current wireless carrier
technology from the gigahertz range into the terahertz (THz) range and beyond.
In fact, while the conventional THz range spans frequencies from $0.1$ to $10$
THz \cite{IEEE-THZ2012}, in this work we consider also the mid infrared (MIR)
and far infrared (FIR) frequency range, considering communication channels up
to $50$ THz.

Both the THz and free-space optical bands have some common
features; for instance, both require highly directional beam
propagation from the source to the detector. However, one of the
most striking and important differences between free-space optical
and the THz bands is the practical aspect of weather impact.
Atmospheric attenuation, due to ambient humidity and water
absorption, limits the maximum link distance achievable at the THz
range. However, in the presence of dust, fog, and atmospheric
turbulence (scintillation), THz wireless links exhibit very little
degradation in performance compared to free-space optical
links~\cite{Vorrius2015,Ma2015,Su2012,Su2012a}. Under
fog conditions, free-space optical links are completely blocked
while THz links exhibit minimal impact. Similar transmission
windows can be exploited at the MIR and FIR ranges, in particular,
between 15 and 34~THz~\cite{IEEE-THZ2016}. A detailed analysis of
the propagation properties of THz signals for wireless
communications in atmosphere can be found, for instance, in
Ref.~\cite{FedericiTHzA,FedericiTHzB,FedericiTHzC}.

An important aspect of THz communication is that of achieving the
highest levels of security possible where secure distances need to
range from $1$~m up to $1$~km. Applications for secure links
include stealthy short distance communications between military
personnel and vehicles (manned or unmanned). Security has been
considered before in terms of THz communication by exploiting
various characteristics and properties of the THz
band~\cite{Federici2013,Federici2010}. Unfortunately, the security
of all such `classical' communication schemes have their limit in
the sense that they can never be unconditionally secure. This
problem can be fixed by quantum key distribution (QKD)
~\cite{PirsAOP2019,Scarani2009,Weedbrook2012}. QKD profits from
the peculiar properties of quantum physics and quantum
information~\cite{Watrous, Hayashi, Holevo-book}, in particular
the no-cloning theorem~\cite{Scarani2005} and the monogamy of
entanglement~\cite{Horodecki2009}, to achieve levels of security
that are not possible using classical cryptography.

Continuous-variable (CV)
QKD~\cite{Braunstein2005,Weedbrook2012,diamanti-REV} has attracted
increasing attention in the last years. This is due to the high
rates achievable that allow one to approach the ultimate limit of
point-to-point private
communication\textbf{~}\cite{PLOB,pirsREV018,SPIE16}, i.e., the
Pirandola-Laurenza-Ottaviani-Banchi (PLOB) bound~\cite{PLOB},
equal to $-\log_{2}(1-T)$ secret bits per transmission over a
bosonic pure-loss channel with transmissivity $T$. Several CV-QKD
protocols have been proposed relying on
one-way~\cite{Grosshans2002,Grosshans2003,Weedbrook2004,Usenko2015},
and two-way
communication~\cite{S.Pirandola2008,ott-PRA2015,Sun2012,Zhang2013},
measurement-device-independence~\cite{Nphot2015} and exploiting
entanglement in the middle \cite{weedbrookENT}. Several
experimental tests \cite{lodewyckPRA07,Jouguet2012,gehring2015}
have shown the capacity of CV-QKD to achieve high-rate secure
communication over metropolitan distances \cite{pirsNAT-reply016}
and beyond \cite{huang-sci-rep-2016}. Recent studies focused on
improving the achievable distance by Gaussian post-selection
\cite{Fiurasek2012,Walk2012}, exploiting noiseless linear
amplifier \cite{Blandino2012}, and also refining the performance
of classical error correction and privacy amplification stages
\cite{Milicevic2018,joseph}. Others works have shown the
feasibility of CV-QKD at wavelengths longer than optical down to
the microwave
regime~\cite{Weedbrook2010,Weedbrook2012a,Weedbrook2014}, also
incorporating finite-size effects~\cite{thermalFS}

In this work, we study the feasibility of CV-QKD at the terahertz
and the extended terahertz range, i.e., from $100$ GHz to $50$
THz. We assume directional beam propagation, and the best
communication windows available that are between $15$ to $34$~THz.
We consider asymptotic key rates under collective Gaussian
attacks~\cite{Gar06,Nav06,Pir08}, which represent the optimal
eavesdropping~\cite{leverrierCOH} after de Finetti
symmetrization~\cite{Renner2005,Ren09}. In particular, we consider
single-mode Gaussian attacks that are optimal~\cite{ott-PRA2017}
in the context of one-way protocols, while two-mode
strategies~\cite{ott-PRA2015} are typically more effective against
two-way quantum communications~\cite{ott-sci-rep-2016} and
measurement-device-independent protocols. In addition to this, we
give a specific description of the hardware for the optical-THz
conversion, inspired by techniques based on the coupling of
microwave and optical fields~\cite{Andrews2014,Zhang2016} to
phonons of a mechanical resonator
\cite{Kerckhoff2013,belacel2017}.

The structure of this paper is the following. In Sec.~\ref{THz-QKD}, we
introduce the communication protocol at THz frequency, we describe the
encoding mechanism and the general aspects of the security analysis. In
Sec.~\ref{PERF}, we study the performances of THz QKD. In Sec.~\ref{hardware},
we describe the hardware implementation of the optical-to-terahertz frequency
conversion. Finally, Sec.~\ref{conclusions} is left to discussions and conclusions.

\section{Terahertz QKD\label{THz-QKD}}

In this section we describe the THz QKD scheme for Gaussian encoding, along
with the corresponding secret key rates. We assume an eavesdropping where the
injected thermal noise matches the amount of trusted thermal noise used by the
encoder~\cite{Weedbrook2010,Weedbrook2012a}.

\subsection{Encoding}\label{sec:encoding}

\begin{figure}[ptb]
\vspace{-1.5cm}
\par
\begin{center}
\includegraphics[width=0.5\textwidth]{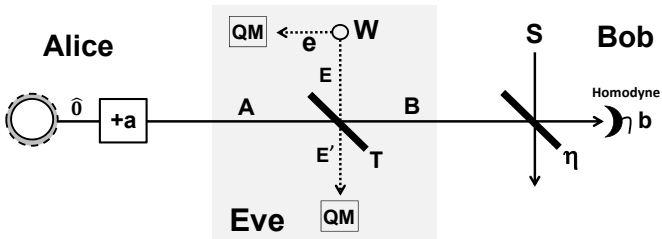}
\end{center}
\par
\vspace{-1.9cm}\caption[protocol]{THz QKD protocol with noisy
(i.e., realistic) homodyne detection. Alice prepares thermal THz
states by displacing their generic quadrature $\hat{0}$ by some
Gaussian variable $a$. She then sends the displaced states through
a thermal-loss channel which is controlled by Eve. This channel
can be described as a beam splitter with transmissivity $T$
subject to thermal noise variance $W$. This noise is produced by
mode $E$ of a two-mode squeezed vacuum (TMSV) state with variance
$W$. Eve's output modes $e$ and $E^{\prime}$ are stored in a
quantum memory (QM) and coherently detected by Eve at the end of
all the communication (collective entangling cloner attack). At
the detection stage, Bob randomly switches the setup of his
homodyne detection, measuring quadrature $q$ or $p$. Bob's output
variable is denoted by $b$. The beam splitter $\eta$ describes the
efficiency of the detector, which is also subject to trusted
thermal noise with variance $S$. We study the performance of the
protocol for several values of the efficiency,
assuming $\eta=10\%$ in a realistic implementation.}%
\label{pic1_new2}%
\end{figure}

The encoding of a protocol for THz QKD is based on the Gaussian
modulation~\cite{Grosshans2002} of thermal states, as also typical
in other studies of QKD at different
wavelengths~\cite{Weedbrook2010,Weedbrook2012a,Weedbrook2014}. In
particular, the protocol studied in this work is based on one-way
communication (see Fig.~\ref{pic1_new2}), and the Gaussian
encoding procedure is the same of that developed in detail in
previous studies of thermal one-way protocol of
Refs.~\cite{Weedbrook2010,Weedbrook2012a}. The sender (Alice)
starts from a vacuum state, $|0\rangle$, and creates coherent
states $|a\rangle$ applying random displacements of the amplitude
$a=q+ i p$. The amplitudes $a$ are chosen from a two-dimensional
Gaussian distribution and contains two independent continuous
variables $q$ and $p$, which will be eventually used to generate a
secret key between Alice and the receiver (Bob).

More in details it goes as follows: Alice randomly displaces the
two quadratures, $q$ and $p$, of a quantum THz source state
(thermal state) according to a bivariate zero-mean Gaussian
distribution. She does this many times, sending each displaced
signal state to the receiver (Bob), over an insecure quantum
channel. The latter can be realistically modeled as a Gaussian
thermal-loss channel~\cite{Weedbrook2012} with transmissivity $T$
and thermal variance $W$.

At the output of the channel, Bob measures the incoming thermal states by
using a THz shot-noise-limited quantum detector; this consists of a noisy
homodyne detection which is randomly switched between the $q$- and the $p$-
quadrature. As depicted in Fig.~\ref{pic1_new2}, Bob's detection can be
described as a beam splitter with transmissivity $\eta$ (mimicking the
detector efficiency) before an ideal homodyne detector. One input port is fed
with the incoming signals, while at the other input is subject to trusted
thermal noise with variance $S$. The detection is followed by steps of
classical post-processing which allows Alice and Bob to extract a
perfectly-correlated bit string. This is the secret key which can be used to
encrypt confidential messages via the one-time pad.

We can describe Alice's mode with the generic quadrature operator $A=\hat
{0}+a$, having total variance $V_{A}=V_{0}+V_{a}$. Here $a$ is a real number
used to denote Alice's classical encoding variable with a variance of $V_{a}$.
Now $\hat{0}$ is the `THz quadrature operator' which can be thought of as the
thermalness of the quantum modes due to the background thermal noise at the
THz frequencies~\cite{Weedbrook2010}. This mode has a corresponding variance
given by $V_{0}$, the `terahertz parameter', defined as $V_{0}:=2\bar{n}+1$,
where $1$ is the vacuum shot noise unit (SNU) and $\bar{n}$ is the mean
thermal photon number. The latter is obtained from Planck's black-body formula
$\bar{n}=[\exp(h\omega/k_{B}t)-1]^{-1}$, where $t$ is the temperature (here
assumed to be $296~$K), $h$ is Planck's constant, $k_{B}$ is Boltzmann's
constant, and $\omega$ is the frequency.

The range of frequencies considered in this work goes from $100$~GHz, for
which we have $V_{0}=123.3$ SNU, to $50$ THz ($V_{0}\simeq1.001$ SNU). Bob's
measurement device extracts the best estimate of Alice's variable $a$.
Specifically, this is achieved by using a shot-noise-limited THz homodyne
detector that randomly switches between the two quadratures. Bob's
corresponding output variable is denoted by $b$. Unlike previous
schemes~\cite{Weedbrook2010,Weedbrook2012a,Weedbrook2014}, here we also take
into account and analyze the effect of a realistic detector at Bob's side.

\subsection{Security of terahertz quantum cryptography}

We will now discuss the eavesdropping strategy against the THz QKD protocol.
First, note that all the elements of the protocol are
Gaussian~\cite{Weedbrook2012}, i.e., the source state is Gaussian and the
detection is Gaussian. It is known that the most powerful attacks against
Gaussian protocols are Gaussian, whose general form has been characterized in
Ref.~\cite{Pir08}. In particular, this attack can be assumed to be collective
under a suitable symmetrization in the limit of large key
length~\cite{Nav06,Gar06,Ren09}. The most important and practical
implementation of this attack is a collective entangling cloner
attack~\cite{Gross03}. This is described in Fig.~\ref{pic1_new2}, where a beam
splitter with transmissivity $T$ simulates the channel attenuation. Then the
thermal noise $W$ is simulated by injecting part of a two-mode squeezed state
(TMSV), which is a zero mean Gaussian two-mode state $\Phi_{eE}(W)$ with
covariance matrix (CM)%
\begin{equation}
\mathbf{V}_{eE}=\left(
\begin{array}
[c]{cc}%
W\mathbf{I} & \sqrt{W^{2}-1}\mathbf{Z}\\
\sqrt{W^{2}-1}\mathbf{Z} & W\mathbf{I}%
\end{array}
\right)  , \label{VeE}%
\end{equation}
where $\mathbf{I}:=\mathrm{diag}(1,1)$ and
$\mathbf{Z}:=\mathrm{diag}(1,-1)$. According to
Fig.~\ref{pic1_new2}, the output ancillas $e$ and $E^{\prime}$ are
stored in a quantum memory and measured by the eavesdropper (Eve)
after Alice and Bob finish the classical communication stage. To
quantify the amount of information gained by Eve, it is then
sufficient to study the total output CM $V_{eE^{\prime}}$ and the
conditional CMs $V_{eE^{\prime}|a}$ and $V_{eE^{\prime}|b}$,
depending on the variable ($a$ or $b$) used for the
reconciliation.

Let us remark a peculiarity of Eve's attack against a thermal
protocol at lower frequencies, compared to a conventional protocol
working at optical or telecom wavelengths ($1550$~nm). In the
optical range, we can set $V_{0}=1$ SNU for the signal states
(coherent states). At frequencies lower than the optical ones, the
shot-noise level of Bob's detector is larger than $1$ SNU and this
is equal to the preparation noise $V_{0}$. For this reason, Eve
may potentially hide herself in this background
noise~\cite{Weedbrook2010,Weedbrook2012a}. For any attenuation $T$
introduced in the communication line, she can compensate for the
reduction of the preparation noise $TV_{0}$ by adding a thermal
noise $W=V_{0}$, so that Bob's detector still gets the same
shot-noise level $TV_{0}+(1-T)W=V_{0}$. Because the value of
$V_{0}$ can become very large at lower frequencies (e.g., below
$1$ THz), this means that Eve may simulate the environmental noise
with a highly-entangled TMSV state $\Phi_{eE}(W)$ which is
therefore very destructive and represents the main bottleneck for
long-distance implementations.

In order to avoid this problem, Bob's detector should be able to
filter the background noise. However, this approach increases the
complexity of the protocol and seems to require the use of
entanglement at the input source. For instance, in quantum
illumination~\cite{Qill0,Qill1,Qill1b,Qill2,Qill3} (see
Ref.~\cite{pirs-sensing} for a recent review), the use of
entangled photons allows the receiver to distinguish the reflected
signal photon (still quantum-correlated with a corresponding idler
photon) from the background of uncorrelated thermal photons.
Thanks to this \textquotedblleft labeling\textquotedblright,
quantum illumination works very well at the microwave frequencies,
therefore providing a basic mechanism for quantum
radar~\cite{Qill3}. This quantum illumination approach was adopted
for private communication in noisy conditions~\cite{Jeff1,Jeff2}
but, as mentioned above, it requires a two-way protocol based on
the use of quantum entanglement at the input.

\subsection{Secret key rates}\label{sec:Secret-key-rates}
In CV-QKD, protocols can be implemented in direct (DR) or reverse
reconciliation configuration (RR). The two approaches have
different security performance, with the RR being more
advantageous to increase the achievable distance over which it is
possible to prepare a secret key.

After the parties have completed the quantum communication (the
distribution of thermal states over the quantum channel), they can
still decide which variable to use to prepare the secret key. The
available data are Alice's classical encoding variable $a$,
described in section \ref{sec:encoding} or Bob's decoding variable
$b$ (see Fig.~\ref{pic1_new2}), i.e., representing the outcomes of
his homodyne detection. If the parties use Alice's encoding
variable ($a$) then the protocol is said to be in
DR~\cite{Grosshans2002}, while using the receiver's decoding
variable ($b$) means that they employ RR~\cite{Grosshans2003}.

The performance of the two reconciliation procedures are not
equivalent. Clearly the correlation between the parties remain the
same, but the accessible information of Eve is reduced in RR. In
fact, while in DR the signals sent are accessible and can be
measured, the outcomes of Bob's detections are not directly
accessible to the eavesdropper, who cannot violate the parties'
private spaces. This advantage for the parties in RR with respect
to the DR, is evident from the fact that security of DR is always
limited to a channel transmissivity $T>0.5$. By contrast, in RR it
is possible to achieve positive key rate also when $T\rightarrow
0$. For each point-to-point CVQKD protocol there are two distinct
key rates to compute, accordingly to the employed reconciliation
strategy.

The security performance of the protocol is assessed by computing
key rate $R$ in the asymptotic regime. The definitions of the key
rate in DR ($\blacktriangleright$) and RR\ ($\blacktriangleleft$),
assuming ideal reconciliation efficiency, are given by the following formulas%
\begin{align}
R^{\blacktriangleright}  &  :=I(a:b)-I(E:a),\label{rateDR}\\
R^{\blacktriangleleft}  &  :=I(a:b)-I(E:b), \label{rateRR-GEN-formula}%
\end{align}
where $I(a:b)$ is Alice and Bob's mutual information, while $I(E:a)$ and
$I(E:b)$ are Eve's accessible information with respect to the variable $a$ and
$b$, respectively. These are bounded by the Holevo
information~\cite{Hol73,Hol99} $\chi^{\blacktriangleright}=\chi(E:a)$ in DR,
and $\chi^{\blacktriangleleft}=\chi(E:b)$ in RR.

Recall that the Holevo information $\chi$ is defined as%
\begin{equation}
\chi:=H-H_{c}, \label{Holevo-GEN}%
\end{equation}
where $H$ $(H_{c})$ is the von Neumann entropy of Eve's total (conditional)
state~\cite{Weedbrook2012}. For a Gaussian state $\rho$, the von Neumann
entropy takes the simple expression%
\begin{equation}
H=\sum_{x}h(x), \label{von Neumann}%
\end{equation}
where $x\geq1$ are the symplectic eigenvalues of the CM\ associated with
$\rho$, and
\begin{equation}
h(x):=\frac{x+1}{2}\log_{2}\frac{x+1}{2}-\frac{x-1}{2}\log_{2}\frac{x-1}{2}.
\label{h-MT}%
\end{equation}

The total von Neumann entropy $H$ is obtained first by computing the CM
$\mathbf{V}_{eE^{\prime}}$ of Eq.~(\ref{eq:V1wayEVE}), corresponding to Eve's
output state $\rho_{eE^{\prime}}$. From $\mathbf{V}_{eE^{\prime}}$ it is
simple to obtain the pair of symplectic eigenvalues (see Appendix~\ref{app1}
for details). In the asymptotic limit of high modulation ($V_{a}\gg1$), the
total von Neumann entropy is given by%
\begin{equation}
H=h\left(  W\right)  +\log_{2}\frac{e}{2}(1-T)V_{a} \label{HT}%
\end{equation}
To study DR we compute the conditional CM $\mathbf{V}_{eE^{\prime}|a}$ from
which one obtains the corresponding von Neumann entropy $H_{c}%
^{\blacktriangleright}$ for Eve's output state conditioned to Alice's
encoding, i.e., $\rho_{eE^{\prime}|a}$. After some algebra, one obtains the
following asymptotic expression%
\begin{equation}
H_{c}^{\blacktriangleright}=h(\bar{\nu}_{1})+\frac{1}{2}\log_{2}\frac{e^{2}%
}{4}(1-T)\Lambda(W,V_{0})V_{a}, \label{HCDR}%
\end{equation}
where
\begin{equation}
\Lambda(x,y):=Tx+(1-T)y,
\end{equation}
and the explicit expression of the eigenvalue $\bar{\nu}_{1}$ is given in
Eq.~(\ref{ni1bar}) of Appendix~\ref{app1}.

For RR we only need to compute the CM\ describing Eve's output
state conditioned to Bob's variable $b$, i.e.,
$\rho_{eE^{\prime}|b}$. This is obtained by completing the CM
$\mathbf{V}_{eE^{\prime}}$ with the entries describing Bob's mode
$b$ and its correlation with Eve's output modes. Then we apply
homodyne detection to Bob's mode $b$, to obtain the conditional CM
$\mathbf{V}_{eE^{\prime}|b}$, whose explicit expression is given
in Eq.~(\ref{VEc}) of Appendix~\ref{app1}. From
$\mathbf{V}_{eE^{\prime}|b}$, we compute the conditional
symplectic spectrum and, taking the asymptotic limit, we derive
the following conditional von Neumann entropy
\begin{align}
H_{c}^{\blacktriangleleft}  &  =h\left(  \bar{\nu}_{3}\right)  +\nonumber\\
&  \frac{1}{2}\log_{2}\frac{e^{2}}{4}\frac{(1-T)\left[  S\left(
1-\eta\right)  \left(  1-T\right)  +W\eta\right]  }{T\eta}V_{a}, \label{HCRR}%
\end{align}
where the explicit expression of the symplectic eigenvalue
$\bar{\nu}_{3}$ can be found in Appendix ~\ref{app1}
.

Using the expression for the Alice and Bob's mutual information%
\begin{equation}
I(a:b)=\frac{1}{2}\log_{2}\frac{\eta TV_{a}}{\eta TV_{0}+\eta(1-T)W+(1-\eta
)S},
\end{equation}
whose derivation is detailed in Appendix \ref{app1}, we obtain the
following key rate in DR
\begin{gather}
R^{\blacktriangleright}(V_{0},T,W,\eta,S)=h\left[  \sqrt{\frac{W\Lambda
(1,WV_{0})}{\Lambda(W,V_{0})}}\right] \label{rate1DR-MT}\\
-h(W)+\frac{1}{2}\log_{2}\frac{T\eta\Lambda(W,V_{0})}{(1-T)[\eta\Lambda
(V_{0},W)+\left(  1-\eta\right)  S]},\nonumber
\end{gather}
and the following one in RR
\begin{gather}
R^{\blacktriangleleft}(V_{0},T,W,\eta,S)=h\left(  \bar{\nu}_{3}\right)
-h\left(  W\right) \label{rateRR-MT}\\
~+\frac{1}{2}\log_{2}\frac{\left(  1-\eta\right)  \left(  1-T\right)  S+W\eta
}{(1-T)[\eta\Lambda(V_{0},W)+(1-\eta)S]}.\nonumber
\end{gather}
The details of the steps to obtain Eq.~(\ref{rate1DR-MT}) and
(\ref{rateRR-MT}) are given in Appendix~\ref{app1}.

\section{Performance in the extended terahertz range\label{PERF}}

In this section, we study the performance of the key rates in DR
and RR. In particular, in subsection~\ref{PERF1}, we plot the
secret-key rates and the security thresholds, while in
subsection~\ref{PERF2} we express the security thresholds in terms
of maximally-achievable frequencies versus channel transmissivity.
We assume a collective entangling-cloner attack where Eve hides in
the detection/preparation noise $W=V_{0}$. Then, in order to
optimize the protocol, one can check that the DR rate in
Eq.~(\ref{rate1DR-MT}) is
maximized by $S=1$, so that we may write%
\begin{align}
R^{\blacktriangleright}(V_{0},T,\eta)  &  =h\left[  \sqrt{T+(1-T)V_{0}^{2}%
}\right]  -h(V_{0})\nonumber\\
&  ~+\frac{1}{2}\log_{2}\frac{T\eta V_{0}}{(1-T)(\eta V_{0}+1-\eta)}.
\label{eqq1}%
\end{align}

For the RR rate in Eq.~(\ref{rateRR-MT}), we consider both the absence of
trusted noise ($S=1$) and its presence by setting $S=V_{0}$. Therefore, our
rate is the maximum
\begin{align}
&  R^{\blacktriangleleft}(V_{0},T,\eta)\nonumber\\
&  :=\max\{R^{\blacktriangleleft}(V_{0},T,V_{0},\eta,1),R^{\blacktriangleleft
}(V_{0},T,V_{0},\eta,V_{0})\},
\end{align}
where%
\begin{gather}
R^{\blacktriangleleft}(V_{0},T,W=V_{0},\eta,S=1)\nonumber\\
=h\left[  \sqrt{\frac{\left[  (1-T)\left(  1-\eta\right)  +\eta\right]  V_{0}%
}{\left(  1-T\right)  \left(  1-\eta\right)  +V_{0}\eta}}\right]  -h\left(
V_{0}\right) \nonumber\\
+\frac{1}{2}\log_{2}\frac{\left(  1-\eta\right)  \left(  1-T\right)
+V_{0}\eta}{(1-T)[\eta V_{0}+1-\eta]}, \label{eqq2}%
\end{gather}
and
\begin{gather}
R^{\blacktriangleleft}(V_{0},T,W=V_{0},\eta,S=V_{0})\nonumber\\
=h\left[  \sqrt{\frac{(1-T)\left(  1-\eta\right)  V_{0}+\eta}{\left(
1-T\right)  \left(  1-\eta\right)  +\eta}}\right]  -h\left(  V_{0}\right)
\nonumber\\
~+\frac{1}{2}\log_{2}\left(1-(1-\eta)T\right)
-\frac{1}{2}\log_{2}(1-T). \label{eqq3}
\end{gather}
Details on the computation of Eq.~(\ref{eqq2}) and (\ref{eqq3})
can be found in Appendix~\ref{app1}.

Now notice that the preparation noise
 $V_{0}$ is uniquely
determined by the frequency $\omega$ (at fixed temperature
$t\simeq296~$K), so that we may set $R=R(\omega,T,\eta)$ in the
previous formulas. Then, solving the equation $R(\omega,T,\eta)=0$
provides the security threshold $\omega=\omega(T,\eta)$ in terms
of minimum-tolerable or accessible frequency versus transmissivity
$T$ and detector efficiency $\eta$. In the following section, we
study the behavior of the rates in DR and RR, finding an optimal
frequency at $\omega_{\text{max}}=30$ THz. We then compare these
optimal rates with the PLOB upper-bound for a thermal-loss
channel~\cite{PLOB}, with transmissivity $T$ and thermal noise
$W=V_{0}(\omega)$ determined by the optimal frequency
$\omega_{\text{max}}=2\bar{n}_{\text{max}}+1$. This upper-bound
takes the analytical form
\begin{equation}
\mathcal{B}=-\log_{2}[(1-T)T^{\bar{n}_{\text{max}}}]-h[V_{0}(\omega
_{\text{max}})], \label{PLOB-TH}%
\end{equation}
for $\bar{n}_{\text{max}}<T(1-T)^{-1}$, while $\mathcal{B}=0$ otherwise.

\begin{figure*}[ptb]
\vspace{-0.0cm}
\par
\begin{center}
\includegraphics[width=0.9\textwidth]{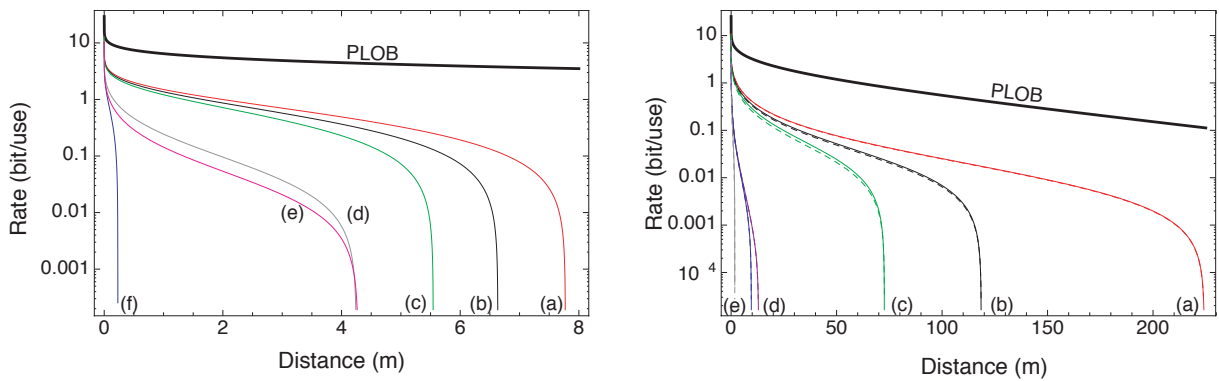}
\end{center}
\par
\vspace{0cm}\caption[protocol]{(Color online) Secret-key rates (bits/use)
versus distance (m) assuming a detection efficiency $\eta=10\%$. The
\textbf{left} panel refers to DR. The curves are as follows: (a) $30~$THz, (b)
$20~$THz, (c) $15$ THz, (d) $100$~GHz, (e) $200$~GHz, and (f) $40$ THz. We
also show the PLOB upper bound $\mathcal{B}$ for a thermal-loss channel which
is computed assuming the parameters of the optimal curve (a). For each curve
we have a specific value of $\delta$~\cite{IEEE-THZ2012,IEEE-THZ2016}, playing
a crucial role in the achievable distance. We have $\delta=50$ dB/Km for
$15-34$ THz, therefore for curves (a), (b) and (c) (and the PLOB bound). For
$\omega$ in the range $40-55$ THz, we have $\delta=1.77\times10^{3}$ dB/Km.
From $100$ GHz to $10$ THz, there are several transmission windows, within a
generally rising atmospheric absorption: At $100$ GHz $\delta=0.6$ dB/Km, at
200 GHz $\delta=1.2$~dB/Km, at $1$ THz $\delta=10^{2}~$dB/Km, while at $10$
THz $\delta=10^{3}$ dB/Km. The \textbf{right} panel refers to RR. The curves
are as follows: (a) $30$~THz, (b) $20$~THz, (c) $15$ THz, (d) $50$~THz (purple
line) and $40$~THz (blue line), (e) $10$~THz. Here too we show the PLOB upper
bound $\mathcal{B}$ for a thermal-loss channel.}%
\label{THzDR}%
\end{figure*}

\subsection{Behavior of the secret key rates\label{PERF1}}

Let us analyze the behavior of the previous rates
$R=R(\omega,T,\eta)$. We assume detection efficiency $\eta=10\%$
and, for several frequencies $\omega$, we study the behavior of
the rate $R$ versus distance $d$ (km) by setting
$T=10^{-\frac{\delta d}{10}}$, where $\delta$ describes the
atmospheric loss (dB/km). As shown in the left panel of
Fig.~\ref{THzDR}, we find that the best performance of the DR rate
$R^{\blacktriangleright}$ is obtained for the frequency range
$15-34$~THz. In particular, the optimal key rate occurs at
$\omega=30$~THz and follows the scaling of the PLOB bound
$\mathcal{B}$ until about $7~$m. This result shows that THz QKD is
possible in DR, with a key rate of $10^{-3}$ (bit per use) over
short distances of order of meters, despite the low detector
efficiency $\eta$. Then, in the right panel of Fig.~\ref{THzDR},
we see that the best performance of the RR rate
$R^{\blacktriangleleft}$ is also achieved at $\omega=30$~THz,
which may allow a secret-key rate of $10^{-4}$ bit/use over a
distance of about $220$~m.

There are two features that mainly deteriorate the performance: one is the
large thermal background that demands to consider entangling cloner attacks
with increasingly large thermal variance $W=V_{0}(\omega)$, as we move to the
low frequency range of the electromagnetic spectrum. The other is the
atmospheric absorption condition that, after starting very low in the
microwave range ($\omega<100$ GHz), grows exponentially as we move towards
higher frequencies. The atmospheric attenuation is caused by the increasing
coupling of the electromagnetic signals with the gases and molecules composing
the atmosphere and may even varying drastically, depending on the weather
conditions~\cite{IEEE-THZ2012,IEEE-THZ2016}.

For the sake of simplicity we assume attenuation under fairly clear
atmospheric conditions noticing that, even in this case, it can vary
considerably from $\delta=0.6$~dB/km at $100$ GHz to $\delta=10^{2}$ dB/km at
$1$ THz to arrive to $\delta=10^{3}$ dB/km at $10$ THz. After this range of
frequencies absorption decreases and several transmission windows become
available, for instance between $15$ and $34$ THz, where the loss rate is
estimated to be around $\delta=50$ dB/km, under pristine
conditions~\cite{IEEE-THZ2016}. Then, the attenuation rises again for
frequencies above this range, but less sharply than before: at $50$~THz we
have again losses as high as approximately $\delta=1.77\times10^{3}$~dB/km.

\subsection{Accessible frequencies\label{PERF2}}

\begin{figure}[ptb]
\vspace{-0.0cm}
\par
\begin{center}
\includegraphics[width=0.45\textwidth]{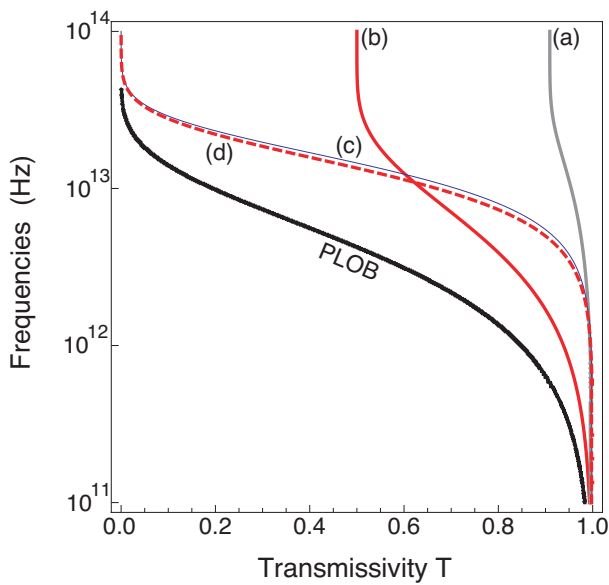}
\end{center}
\par
\vspace{0cm}\caption[protocol]{(Color online) Security thresholds
in terms of accessible frequencies $\omega$ (Hz) versus channel
transmissivity $T$. The range of values of $T$ and $\omega$, for
which for which the protocol is secure, are those \emph{above} the
curves. The curves are: (a) DR with $\eta=10\%$, (b) DR with
$\eta=100\%$, (c, solid blue) RR with $\eta=100\%$, (d, dashed
red) RR with $\eta=10\%$ and detector's trusted noise $S=V_{0}$.
We compare these results with the threshold associated with the
PLOB bound of
Eq~(\ref{PLOBsec}).}%
\label{fig-freq}%
\end{figure}

Let us now study the security threshold expressed in terms of accessible
frequencies versus channel transmissivity and detector efficiency, i.e.,
$\omega=\omega(T,\eta)$. This analysis follows and extends a similar approach
described in Refs.~\cite{Weedbrook2010,Weedbrook2012a,Weedbrook2014}. We show
the performance in DR and RR, and we also compare it with the threshold coming
from the PLOB upper bound $\mathcal{B}=0$, which is equal to $\bar
{n}=T(1-T)^{-1}$, i.e.,
\begin{equation}
\omega=\frac{1+T}{1-T}. \label{PLOBsec}%
\end{equation}
In Fig.~\ref{fig-freq} positive rates are those \textit{above }the thresholds.
As we can see there is a window of positive rates opening already at $1$ THz
according to the PLOB bound. However, our protocols start to work well (in
both DR and RR) from about $10$~THz, where wide ranges of transmissivities are
accessible.

In particular, note that the RR security threshold $\omega^{\blacktriangleleft
}=\omega^{\blacktriangleleft}(T,\eta)$, coming from $R^{\blacktriangleleft}%
=0$, is computed assuming detection efficiency $\eta=10\%$. This efficiency
can be obtained as a two-step operation by combining a preliminary conversion
from the THz frequencies to optical, followed by an optical homodyne
detection, which can be performed with nearly unit efficiency \cite{hom1,hom2}%
. Therefore the overall efficiency $\eta$ is limited by the efficiency
$\tilde{\eta}$ of the THz-optical conversion. Recent
works~\cite{Andrews2014,belacel2017} suggests that it is reasonable to set
this efficiency $\tilde{\eta}=10\%$. In the following section, we describe a
possible hardware realization for a coherent THz-optical converter.

\section{Physical hardware realization\label{hardware}}
\subsection{THz-optical converter}
\begin{figure}[t]
\vspace{-0.0cm}
\par
\begin{center}
\includegraphics[width=0.475\textwidth]{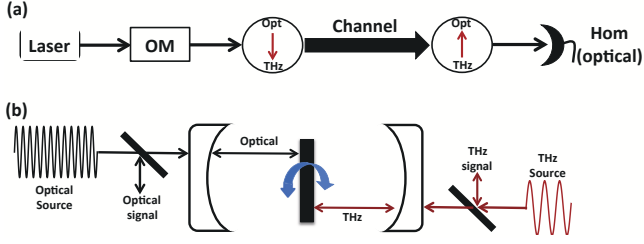}
\end{center}
\par
\vspace{0cm}\caption[protocol]{(Color online) Potential hardware
implementation of a THz-optical converter. (a) Optical signals generated by a
laser are optically modulated (OM) and then converted from optical frequencies
to THz. The THz signals are sent through the channel, and their outputs are
converted back to optical. Finally, Bob performs an optical homodyne detection
(hom). Note that the attenuation of the first optical-to-THz conversion can be
mitigated by increasing the input power level before the converter. (b) We
show how a THz-optical converter could work, exploiting the mediation via a
common mechanical mode. The cavities are schematically represented as
Fabry-Perot, but also implementation based on co-locating optical, THz and
phonon modes in photonic/phononic crystals could be considered. Co-location of
optical and phonon modes has been proved in Ref.~\cite{kakimi014}, as well as
photonic crystals implementation~\cite{eich2009}.}%
\label{HTSHES}%
\end{figure}

Terahertz sources, modulators and detectors (see Fig.~\ref{HTSHES} (a)) are
substantially less well-developed than their optical counterparts. Here we
sketch a hardware realization of a terahertz quantum communication system
based on coherent terahertz-optics conversion and existing optical technology.
In particular, we describe the operation of a coherent terahertz-optical
interface, mediated via a low-frequency mechanical resonance. This approach
for THz QKD is motivated by recent progress in the development of coherent
microwave-optical interfaces~\cite{Bochmann2013,Bagci12014,Andrews2014} which
can be used to create a hybrid quantum
network~\cite{Kimble2008,pirs-sam-NAT-2016,TeleREV}. Such interfaces operate
via the coupling of microwave and optical fields to a common phononic mode in
a solid-state material.

The proposed bidirectional terahertz-optical converter (see Fig.~\ref{HTSHES}
(b)) consists of a THz and optical cavity mode, each coupled to a common
(linearized) phononic mode, and out-coupled to a THz and optical waveguide,
respectively. The phononic mode is sideband-coupled to both the THz and the
optical cavity modes. The terahertz and optical cavity mode resonance
frequencies are denoted by $\omega_{t}$ and $\omega_{0}$, and the decay rates
into the coupled waveguide modes are denoted by $\kappa_{t}$ and $\kappa_{0}$,
respectively. The phononic mode frequency is denoted by $\omega_{m}$, and its
effective coupling rates to the terahertz and optical cavity modes are given
by $g_{t}$ and $g_{0}$, respectively.

This device may be analyzed using the input-output theory of quantum
optics~\cite{gardiner}, as described in detail in
Appendix~\ref{app:converteranalysis}. Denoting the quadrature frequency
components of the optical input and output fields by $q_{o,in/out}[\omega]$
and of the terahertz input and output fields by $q_{t,in/out}[\omega]$, the
terahertz-optical and optical-terahertz frequency-dependent transmissivity may
be defined by
\begin{equation}
t(\omega)=\frac{\langle q_{o,out}(\omega)\rangle}{\langle q_{t,in}%
(\omega)\rangle}=\frac{\langle q_{t,out}(\omega)\rangle}{\langle
q_{o,in}(\omega)\rangle}, \label{eq:tMainText}%
\end{equation}
where $\omega$ is defined with respect to $\omega_{t}$ $(\omega_{o})$ for the
terahertz (optical) mode. It may be shown that the zero-frequency
transmissivity is given by
\begin{equation}
|t(\omega=0)|=\frac{8g_{o}g_{t}\sqrt{\kappa_{o}\kappa_{t}}}{4(g_{o}^{2}%
\kappa_{t}+g_{t}^{2}\kappa_{o})+\kappa_{o}\kappa_{m}\kappa_{t}}.
\label{eq:tomega0MainText}%
\end{equation}
This transmissivity goes to one in the limit of negligible damping $\kappa
_{m}\rightarrow0$, symmetric couplings $g_{o}=g_{t}=g$, and symmetric
out-couplings $\kappa_{o}=\kappa_{t}=\kappa$.

Now for high-fidelity conversion, in addition to a near-unity zero-frequency
gain, the converter must also have a flat magnitude response and a linear
phase response over a reasonable bandwidth. As detailed in
Appendix~\ref{app:converteranalysis}, this would be the case over a range of
frequencies $\omega$ such that
\begin{equation}
|\omega|\ll g,\kappa,4g^{2}/\kappa. \label{eq:BWMainText}%
\end{equation}
This might lead to a useful modulation bandwidth of the order of
$\sim10\,\mathrm{MHz}$, sufficient for the required coherent field modulation.

\subsection{Analysis of the noise from the optical-to-THz conversion, and
impact on the performance \label{noise-conv}}

In this section we study the impact of the noise coming from the
optical-to-THz converters, on the performance of our THz QKD
scheme. We consider the protocol used in RR and, particularly, we
focus on the three cases showing the best performance under ideal
conditions: $30$ THz (a), $20$ THz (b), and $15$ THz (c) (see
Fig.~\ref{THzDR}) (right-panel).
\begin{figure}[t] \vspace{0.5cm}
\par
\begin{center}
\includegraphics[width=0.45\textwidth]{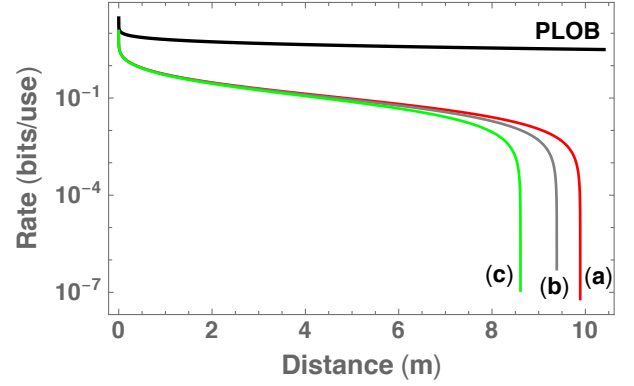}
\end{center}
\par
\vspace{0cm}\caption[protocol]{(Color online) Secret-key rates
(bits/use) versus distance (m) for the protocol in RR, including
the noise from the frequency conversion process. The curves refer
to cases $30~$THz (a), $20~$THz (b), $15$ THz (c). The black solid
line describes the PLOB upper bound $\mathcal{B}$ for a
thermal-loss channel which is computed assuming the atmospheric
attenuation $\delta=50$ dB/Km for frequencies in the transmission
window of $15-34$ THz \cite{IEEE-THZ2012,IEEE-THZ2016}. We assume
cryogenic temperature of $1$K. In this conditions the conversion
process is still noisy, with thermal occupation number
$\bar{n}_{THz}=1.72$ photons populating the THz cavity-mode. The
presence of such noise reduces the achievable distance: At $15$
THz the achievable distance is $12$ m while, despite case (a)
($30$ THz) remains the best of the three, it is also the most
affected from the presence of this additional noise, with the
achievable distance passing from $225$ m to
$13.5$ m.}%
\label{RR-THz-CONV-noise}%
\end{figure}

The noise affecting the conversion process is quantified
calculating the effective thermal occupations number of the THz
and optical modes inside the cavity. To perform this calculation,
we assume to work within the linear approximation, obtaining the
steady-state solving Lyapunov equation \cite{MA}. We then compute
the steady-state correlation functions, via the quantum regression
theorem. The computation of the spectra of the optical and THz
modes, is obtained computing the Fourier transforms of the
correlation functions and, integrating the spectra, we obtain the
effective thermal occupation numbers of the cavity modes.

To minimize the presence of unwanted thermal photons during the
frequency-conversion process, the converter described in
Sec.~\ref{hardware} (see also Fig.~\ref{HTSHES}) is assumed to
operate at cryogenic temperature. For the three cases, $30$ THz
(a), $20$ THz (b) and $15$ THz (c), we find that at $1$K the
population of the cavity modes is dominated by thermal photons in
the THz mode. The thermal occupation number is given by
$\bar{n}_{THz}=1.72$ photons, and this number remains the same for
cases (a), (b) and (c). Consequently, the thermal noise caused by
the presence of these photons, $V_{c}=2\bar{n}_{THz}$, represents
additional thermal noise assumed under Eve's control, and its
presence clearly degrades the performance.

The results of this analysis are shown in
Fig.~\ref{RR-THz-CONV-noise}, where we plot the three key-rates
(bit/use) for cases (a), (b), and (c). The noise of the converter
may have a strong impact on the performance, reducing the
achievable distance by one order of magnitude, with respect the
ideal case described by Fig.~\ref{THzDR}) (right-panel). We remark
that the solution proposed in this section does not represent the
optimal solution. To keep the performance closer to the ideal
conditions described in Fig.~\ref{THzDR}), further studies are
required to reduce the noise in the converters or, avoid the use
of converters, improving the performance of homodyne detection at
frequencies alternative to the optical one.

\section{Discussion and Conclusion\label{conclusions}}

In this work, we have studied the feasibility of quantum cryptography working
in the THz range of frequency of the electromagnetic spectrum, going from
$0.1$ to $50$ THz. This frequency range is attractive for the potential
boosting of data rate of wireless communication. To perform this study we
assumed unidirectional, plane-wave emission setup, and we performed the
security analysis in the asymptotic limit of many uses of the quantum channel,
bounding the eavesdropper's accessible information by the Holevo bound. We
focused on a switching protocol, where the receiver decoding is performed by
randomly switching between the two possible quadratures to homodyne. We
assumed detections performed in the THz range with an efficiency of $10\%$,
which represents the performance of recent optical-to-microwave conversion
rate and state-of-the-art THz
devices~\cite{Bochmann2013,Bagci12014,Andrews2014,belacel2017,kakimi014}.

We found that QKD at THz frequencies is possible in both direct and reverse
reconciliation, over short distances, because of the following trade-off: THz
radiation is strongly affected by atmospheric loss, caused by suspended
molecules in atmosphere, which sharply increase losses as we move away from
the microwave regime. By contrast moving towards the microwave regime, we have
to deal with a large thermal background. This second aspect is particularly
harmful for signals below $1$ THz, where the thermal background is already
more than one order-of-magnitude higher than the vacuum noise. The
eavesdropper can exploit this high level of shot noise for her attack.

The range of distances over which it is possible to implement THz
QKD is indeed short, if compared to optical communication. In
particular, we found that in direct reconciliation we can achieve
a key rate of $10^{-2}$ bit/use over a distance of meters, about
$4$~m at $0.1$~THz and $7.5~$m at $30$ THz. This can be used for
short-range applications, such as credit card/ATM transactions or
proximity and access cards, e.g., for secure access of buildings.
Then, using reverse reconciliation at $30$~THz, we obtain a key
rate of $10^{-3}$ bit/use, over a distance of about $220$~m.
Finally, we have considered the impact of the additional noise,
caused by imperfect optical-THz conversion, on the performance of
the scheme. We focused on the protocol in reverse reconciliation,
determining the thermal noise coming from the cavity modes at the
cryogenic temperature of $1$K. The key rates for the three optimal
cases, $15$, $20$ and $30$ THz are plotted in
Fig.~\ref{RR-THz-CONV-noise}. The presence of unwanted thermal
photons, from a noisy conversion, increases the total noise and
degrades the performance of the scheme. The achievable distance
still remains useful for short range communications, for instance
within buildings (see Fig.~\ref{RR-THz-CONV-noise}). The
extra-noise in the converter is caused by the sensitivity to
thermal noise of the intermediate quasiparticle mode. A possible
solution to mitigate this aspect, may be to implement a converter
scheme insensitive to this type of noise on the intermediate
quasiparticle mode, as discussed in ref.~\cite{WANG}. Other
solutions may require the development of novel designs, able to
overcome the imperfections of transduction schemes and improve
performances \cite{ClerknpjQI2019}.

The analysis of the security thresholds at various frequencies,
summarized in Fig.~\ref{fig-freq}, shows that direct
reconciliation is limited to high-transmission conditions, mainly
as a consequence of the low-efficiency of $10\%$ of Bob's
detector. By contrast, reverse reconciliation is less affected by
the low detector efficiency and it is possible to obtain a
positive key at transmissivity as low as $T\simeq0.1$. In addition
to this, the use of trusted-noise-assisted detection helps to
increase the security threshold, despite the fact that the benefit
is incremental.

Regarding the possibility of achieving long-distance THz quantum
cryptographic communication with our scheme, let us stress that
the limited achievable distance emerged from our analysis is
mainly consequence of atmospheric absorption (see parameters used
in Fig.~\ref{THzDR}). Removing the limitation our scheme can be
much more effective. Possible application can be in Space
satellite-to-satellite communication, where attenuation is caused
by geometrical losses \cite{PirsAOP2019}, caused by the
diffraction of the Gaussian beam traveling from the sender to the
receiver. In such a case our scheme gives much better performance
in terms of achievable distance. For instance,
satellite-to-satellite THz QKD, over the range of hundreds of Km,
represents an achievable target, with the distance depending on
the frequency employed, and also the operative temperature of the
devices.

In conclusion, we have explored the possibility of implementing
quantum cryptography in the THz regime, finding that this is
possible over short distances with suitably high rates. The
distance range depends crucially on the transmission windows
considered. Further analysis may take into account the impact of
finite-size effects, and the possibility of using other
communication protocols, like two-way schemes. Let us also remark
that working in the THz range of frequencies can also provide some
technical advantage with respect to the standard optical regime.
At optical frequencies, an intense oscillator signal is typically
used to keep synchronized the phases of Alice's and Bob's systems.
By contrast, at much lower frequencies as the THz ones, we may use
local clocks to maintain the relative phase of the quadratures.

\section{Acknowledgements}

We thank Michael Leuenberger for discussions. This work has been
supported by the EPSRC via the `UK quantum communications hub'
(Grant no. EP/M013472/1), and the European Commission under the
project CiViQ (Grant no. 820466).

\appendix

\section{Details of the computation of the secret-key rates\label{app1}}

\subsection{Mutual information}

We begin the security analysis by noting that, at Bob's side, the
loss and noise affecting his detector can be modeled by a beam
splitter with transmission $\eta$ processing the incoming mode
with a trusted mode, with variance $S$, being one of the two
component of  two-mode squeezed vacuum (TMSV) state, described by
a CM of the following form
\begin{equation}
\mathbf{V}_{S}=\left(
\begin{array}
[c]{cc}
S\mathbf{I} & \sqrt{S^{2}-1}\mathbf{Z}\\
\sqrt{S^{2}-1}\mathbf{Z} & S\mathbf{I}.
\end{array}
\right)
\end{equation}

To assess the security of a QKD protocol, we start quantifying the
correlation between the parties, i.e., the first quantity to
calculate is the mutual information between Alice and Bob
$I(a:b)$, which is identical for both DR and RR. The mutual
information is defined as \cite{PirsAOP2019}
\begin{equation}
I(a:b)=\frac{1}{2}\log_{2}\frac{V_{b}}{V(b|a)} \label{IAB-GEN}
\end{equation}
where $V_{b}$ is the variance of Bob's output mode $b$ and is given by
\begin{equation}
V_{b}=\eta TV_{A}+\eta(1-T)W+(1-\eta)S, \label{Vb}
\end{equation}
with $V_{A}=V_{0}+V_{a}$. The conditional variance can be obtained
from Eq.~(\ref{Vb}) setting the classical Gaussian variance
$V_{a}=0$,
\begin{equation}
V(b|a)=\eta TV_{0}+\eta(1-T)W+(1-\eta)S. \label{Vba}
\end{equation}
It describes the probability of getting the outcome $a$ measuring
the incoming signals. Now, using Eqs.~(\ref{Vb}) and (\ref{Vba}),
we may write Alice and Bob's mutual information
\begin{equation}
I(a:b)=\frac{1}{2}\log_{2}\frac{\eta
TV_{A}+\eta(1-T)W+(1-\eta)S}{\eta TV_{0}+\eta(1-T)W+(1-\eta)S},
\label{eq:Iab}
\end{equation}
and, taking the (ideal) limit of asymptotically large signal
modulation $V_{a}\gg1$, Eq.~(\ref{eq:Iab}) becomes
\begin{equation}
I(a:b)\rightarrow\frac{1}{2}\log_{2}\frac{\eta TV_{a}}{\eta TV_{0}%
+\eta(1-T)W+(1-\eta)S}. \label{eq:Iab-asy}
\end{equation}
We remark that the asymptotic approximation represents a solid
estimation of security and performance of a Gaussian protocol
\cite{PirsAOP2019}. However, in case of practical implementations,
additional finite-size and composable security analysis will be
needed.

\subsection{Direct Reconciliation}

In section \ref{sec:Secret-key-rates} we discussed the two
possible implementation of CV-QKD protocols. In the following
sections we discuss the details of the computation of the key
rates in direct and reverse reconciliation.

We assume the Eve's dropper starting from a TMSV state, composed
by her ancillary modes $e$ and $E$. She keeps $e$ and send $E$
through the channel, coupling it with the incoming signals. It is
simple to compute the CM corresponding to Eve's output modes $e$
and $E^{\prime}$, which is given by
\begin{equation}
\mathbf{V}_{eE^{\prime}}=\left(
\begin{array}
[c]{cc}%
W\mathbf{I} & \sqrt{T(W^{2}-1)}\mathbf{Z}\\
\sqrt{T(W^{2}-1)}\mathbf{Z} & TW+(1-T)V_{A}\mathbf{I}
\end{array}
\right)  ~. \label{eq:V1wayEVE}
\end{equation}
To compute Eve's accessible information, we need to perform the
symplectic analysis \cite{Weedbrook2012} and obtain the symplectic
eigenvalues. These are then used to compute the Holevo function,
which bounds Eve's accessible information in case of collective
attacks \cite{PirsAOP2019}.

This analysis is simplified in the limit of large modulation
($V_{a}\gg1$): First we compute the symplectic spectrum of the CM
in Eq.~(\ref{eq:V1wayEVE}), by finding the standard eigenvalues of
$\nu=|i\Omega V_{b}|$ ~\cite{Weedbrook2012}, and then we take the
limit of large $V_{a}$. It is easy to check that one obtains the
following pair of distinct symplectic eigenvalues
\begin{align}
\nu_{1}  &  \rightarrow W\label{ni1}\\
\nu_{2}  &  \rightarrow(1-T)V_{a}. \label{ni2}%
\end{align}
Consequently, we can compute Eve's total entropy
$H=\sum_{k}h(\nu_{k})$, where $\nu_{k}$ are symplectic
eigenvalues. We find that it is given by
\begin{equation}
H=h(\nu_{1})+h(\nu_{2})\rightarrow h(W)+\log\frac{e}{2}(1-T)V_{a},
\label{eq:Se1wayDR}
\end{equation}
where we have used the function $h(x)$
\begin{equation}
h(x):=\frac{x+1}{2}\log_{2}\frac{x+1}{2}-\frac{x-1}{2}\log\frac{x-1}{2},
\end{equation}
and its asymptotic limit $h(x)\rightarrow\log\frac{e}{2}x$ for
$x\rightarrow\infty$~\cite{Weedbrook2012}.

Similarly to the computation for the conditional variance of
Eq.~(\ref{Vba}) Eve's conditional CM $\mathbf{V}_{eE^{\prime}|a}$
can be computed by setting $V_{a}=0$ in
$\mathbf{{V}}_{eE^{\prime}}$, of Eq.~(\ref{eq:V1wayEVE}), for one
of the two quadratures only, because the receiver is collecting
outcomes only from one of the two quadratures. Adopting the same
steps described above to perform the symplectic analysis, the
asymptotic conditional symplectic spectrum is given by
\begin{align}
\bar{\nu}_{1}  &  \rightarrow\sqrt{\frac{W\Lambda(1,WV_{0})}{\Lambda(W,V_{0}%
)}},\label{ni1bar}\\
\bar{\nu}_{2}  &  \rightarrow\sqrt{(1-T)\Lambda(W,V_{0})V_{a}}~,
\label{ni2bar}%
\end{align}
where we have defined the function $\Lambda(x,y):=Tx+(1-T)y$. We
can now compute Eve's conditional entropy, which is therefore
given by
\begin{equation}
H_{c}^{\blacktriangleright}\rightarrow h(\bar{\nu}_{1})+\log\frac{e}{2}%
\sqrt{(1-T)\left[  TW+(1-T)V_{0}\right]  V_{a}}~. \label{eq:Sea1wayDR}%
\end{equation}

Using Eqs.~(\ref{eq:Se1wayDR}) and~(\ref{eq:Sea1wayDR}), Eve's Holevo
information is given by
\begin{align}
\chi(E:a)  &  =H-H_{c}^{\blacktriangleright}\nonumber\\
&  \rightarrow h(W)-h(\bar{\nu}_{1})+\frac{1}{2}\log\frac{(1-T)V_{a}}%
{\Lambda(W,V_{0})}~. \label{eq:Xe1wayDR}%
\end{align}
By using Eq.~(\ref{eq:Xe1wayDR}) and Eq.~(\ref{eq:Iab-asy}) we can
write the final key rate as
\begin{align}
R^{\blacktriangleright}(V_{0},T,W,\eta,S)  &  =\frac{1}{2}\log\frac
{T\eta\Lambda(W,V_{0})}{(1-T)[\eta\Lambda(V_{0},W)+\left(  1-\eta\right)
S]}\nonumber\\
&  +h\left[  \sqrt{\frac{W\Lambda(1,WV_{0})}{\Lambda(W,V_{0})}}\right]
-h(W)~, \label{rate1DR}%
\end{align}
which is given in Eq.~(\ref{rate1DR-MT}) of the main text.

\subsection{Reverse Reconciliation\label{RR-comp}}

Note that we can write Eve and Bob's joint CM as follows
\begin{equation}
\mathbf{V}_{eE^{\prime}B}=\left(
\begin{array}
[c]{cc}%
\mathbf{V}_{E} & \mathbf{C}\\
\mathbf{C}^{T} & \mathbf{B}%
\end{array}
\right)  ~, \label{VEB}%
\end{equation}
where
\begin{equation}
\mathbf{B}=\eta\lbrack\Lambda(V_{0},W)+TV_{A}+((1-\eta)/\eta)S]\mathbf{I},
\end{equation}
and%
\begin{equation}
\mathbf{C}=\left(
\begin{array}
[c]{c}%
\sqrt{\eta(1-T)(W^{2}-1)}\mathbf{Z}\\
\sqrt{\eta T(1-T)}(W-V_{A})\mathbf{I}%
\end{array}
\right)  ~.
\end{equation}

Because Bob uses homodyne detection, Eve's conditional CM is
obtained applying the standard formula for homodyne (Gaussian)
measurements \cite{Weedbrook2012} to the Eve-Bob joint CM of
Eq.~(\ref{VEB}), i.e., that is given by
\begin{equation}
\mathbf{V}_{eE^{\prime}|\beta}=\mathbf{V}_{eE^{\prime}}-\mathbf{C}%
(\boldsymbol{\Pi}\mathbf{B}\boldsymbol{\Pi})^{-1}\mathbf{C}^{T}, \label{VEc}%
\end{equation}
where $\boldsymbol{\Pi}:=\mathrm{diag}(1,0)$%
~\cite{Eisert2002,Fiurasek2002,Weedbrook2012,Gae}. After some
algebra, we find that Eve's conditional covariance matrix is

\begin{equation}
\mathbf{V}_{eE^{\prime}|b}=\left(
\begin{array}
[c]{cc}%
\mathbf{A} & \mathbf{D}\\
\mathbf{D}^{T} & \mathbf{C}%
\end{array}
\right)  ,
\end{equation}
with%
\begin{align}
\mathbf{A}  &  \mathbf{=}W\mathbf{I}-\frac{(1-T)(W^{2}-1)}{\theta}%
\mathbf{\Pi,}\\
\mathbf{C}  &  =\Lambda\left(  W,V_{A}\right)  \mathbf{I}-\frac{\alpha}%
{\theta}\mathbf{\Pi},\\
\mathbf{D}  &  =\sqrt{T(W^{2}-1)}\mathbf{Z}-\frac{\gamma}{\theta}\mathbf{\Pi},
\end{align}
where we defined
\begin{align}
\alpha &  =\left(  1-T\right)  T\left[  W-V_{a}+V_{0}\right]  ^{2},\\
\theta &  :=\Lambda\left(  V_{A},W\right)  +\frac{1-\eta}{\eta}S,\\
\gamma &  :=\left(  1-T\right)  \left(  W-V_{A}\right)  \sqrt{T(W^{2}-1)}.
\end{align}

We then compute the conditional symplectic spectrum. In the limit
of large modulation $(V_{a}\gg1)$, we derive
\begin{align}
\bar{\nu}_{3}  &  \rightarrow\sqrt{\frac{W\left[  S(1-T)\left(
1-\eta\right) +\eta\right]  }{S\left(  1-T\right)  \left(
1-\eta\right)  +W\eta}
},\label{niCRR1}\\
\bar{\nu}_{4}  &  \rightarrow\sqrt{\frac{(1-T)\left[  S\left(
1-\eta\right) \left(  1-T\right)  +W\eta\right]  }{T\eta}V_{a}}.
\label{niCRR2}
\end{align}
These can be used to compute Eve's conditional entropy $H_{c}
^{\blacktriangleleft}$. Therefore, we find
\begin{align}
\chi(E  &  :b)=H-H_{c}^{\blacktriangleleft},\\
&  \rightarrow h(W)-h(\bar{\nu}_{3})+\frac{1}{2}\log_{2}\frac{(1-T)T\eta
V_{a}}{S(1-\eta)(1-T)+W\eta}.\nonumber
\end{align}
Finally, by using $I(a:b)$ previously computed, we can derive the
asymptotic secret-key rate in RR
\begin{align}
R^{\blacktriangleleft}(V_{0},T,W,\eta,S)  &  =h\left(  \bar{\nu}_{3}\right)
-h\left(  W\right)  +\label{rateRR}\\
&  \frac{1}{2}\log\frac{\left(  1-\eta\right)  \left(  1-T\right)  S+W\eta
}{(1-T)[\eta\Lambda(V_{0},W)+(1-\eta)S]}.\nonumber
\end{align}

From Eq.~(\ref{rateRR}) we can compute the explicit expressions of
Eq.~(\ref{eqq2}) and (\ref{eqq3}) in the main text, describing the
key rate in RR for specific values of the trusted noise $S=1$ and
$S=V_{0}$. Using the expression of the symplectic eigenvalue
$\bar{\nu}_{3}$ of Eq.~(\ref{niCRR1}), setting thermal noise
$W=V_{0}$ and the trusted noise $S=1$, from Eq.~(\ref{rateRR}) we
obtain the following expression of the key rate
\begin{eqnarray}
R^{\blacktriangleleft}_{S=1}&=&
h\left[\sqrt{\frac{\left[(1-T)\left(1-\eta\right) +\eta\right]
V_{0}}{\left(1-T\right)\left(1-\eta\right) +V_{0}\eta}}\right]
-h\left(V_{0}\right)\nonumber\\
&&+\frac{1}{2}\log_{2}\frac{\left(1-\eta\right)\left(1-T\right)
+V_{0}\eta}{(1-T)[\eta V_{0}+1-\eta]},\label{eqq2-APP}
\end{eqnarray}
which corresponds to Eq.~(\ref{eqq2}) in the main text.

Similarly, for $S=V_{0}$, the symplectic eigenvalue of
Eq.~(\ref{niCRR1}) is given by
\begin{equation}
\bar{\nu}_{3}(V_{0})=\sqrt{\frac{\left[V_{0}(1-T)\left(1-\eta\right)+\eta
\right]}{\left(1-T\right)\left(1-\eta\right)+\eta}},
\end{equation}
and Eq.~(\ref{rateRR}) can be written as follows
\begin{eqnarray}
R^{\blacktriangleleft}_{S=V_{0}}&=&
h\left[\bar{\nu}_{3}(V_{0})\right]-h\left(V_{0}\right)
\nonumber\\
&&+\frac{1}{2}\log_{2}\frac{\left(1-\eta\right)\left(1-T\right)V_{0}
+V_{0}\eta}{(1-T)[\eta
V_{0}+\left(1-\eta\right)V_{0}]},\nonumber\\
&=& h\left[\bar{\nu}_{3}(V_{0})\right]-h\left(V_{0}\right)
\nonumber\\
&&+\frac{1}{2}\log_{2}\frac{\left(1-\eta\right)\left(1-T\right)
+\eta}{1-T},\nonumber\\
&=&
h\left[\bar{\nu}_{3}(V_{0})\right]-h\left(V_{0}\right)+\frac{1}{2}\log_{2}\left[1-\left(1-\eta\right)T\right]
\nonumber\\
&&-\frac{1}{2}\log_{2}\left(1-T\right),\label{eqq3-APP}
\end{eqnarray}
which corresponds to Eq.~(\ref{eqq3}) in the main text. These
equations describe the key rates in RR, for two distinct values of
the trusted noise $S$, affecting the detection at Bob's side.

\section{Coherent Bidirectional Terahertz-Optical Converter}

\label{app:converteranalysis}

The proposed converter consists of a terahertz cavity mode, an optical cavity
mode, and a phononic mode treated in a linearized approximation. Introducing
the vector of quadrature operators
\begin{equation}
\vec{x}=\left(  q_{o},q_{t},q_{m},p_{o},p_{t},p_{qp}\right)  ^{\mathrm{T}},
\end{equation}
the Hamiltonian, defined in a frame rotating with respect to the oscillator
resonance frequencies, is $H=\vec{x}^{T}\mathbf{G}\vec{x}/2$ where
$\mathbf{G}=\mathrm{Diag}(\mathbf{g},\mathbf{g})$ with
\begin{equation}
\mathbf{g}=\left(
\begin{array}
[c]{ccc}%
0 & 0 & g_{o}\\
0 & 0 & g_{t}\\
g_{o} & g_{t} & 0
\end{array}
\right)  .
\end{equation}
The Heisenberg-Langevin equations corresponding to this Hamiltonian are
\begin{equation}
\dot{\vec{x}}=\mathbf{M}\,\vec{x}+\mathbf{N}\,\vec{x}_{\mathrm{in}},
\label{eq:HLeqns}%
\end{equation}
with the input noise operators are defined in the usual way \cite{gardiner}
and the matrices as
\begin{subequations}
\begin{align}
\mathbf{M}  &  =\left(
\begin{array}
[c]{cc}%
-\mathbf{n}^{2}/2 & \mathbf{g}\\
-\mathbf{g} & -\mathbf{n}^{2}/2
\end{array}
\right)  ,\\
\mathbf{N}  &  =\mathrm{Diag}(\mathbf{n},\mathbf{n}),
\end{align}
where $\mathbf{n}=\mathrm{Diag}(\sqrt{\kappa_{o}},\sqrt{\kappa_{t}}%
,\sqrt{\kappa_{qp}})$.

Taking the Fourier transform of Eq.~(\ref{eq:HLeqns}) and rearranging the
resulting equation, we obtain the Fourier transform of the intracavity
quadrature operators in terms of the input field quadrature operators as
\end{subequations}
\begin{equation}
(\mathbf{M}+i\omega\mathbf{1})\vec{x}(\omega)=-\mathbf{N}\,\vec{x}_{in}%
(\omega). \label{eq:FTofHLeqns}
\end{equation}
Substituting the standard input-output boundary condition \cite{gardiner}, in
the form
\begin{equation}
\vec{x}_{in}(\omega)+\vec{x}_{out}(\omega)=\mathbf{N}\,\vec{x}(\omega),
\end{equation}
into Eq.~(\ref{eq:FTofHLeqns}), and again rearranging, we obtain the output
fields in terms of the input fields as
\begin{equation}
\vec{x}_{out}(\omega)=-\left[
\mathbf{N}(\mathbf{M}+i\omega\mathbf{1}
)^{-1}\mathbf{N}+\mathbf{1}\right]  \,\vec{x}_{in}(\omega).
\end{equation}
Taking the expectation values of this equation, we obtain
\begin{equation}
\langle\vec{x}_{out}(\omega)\rangle=\mathbf{H}(\omega)\langle\vec{x}
_{in}(\omega)\rangle,
\end{equation}
where $\mathbf{H}(\omega)$ is the matrix of frequency responses,
\begin{equation}
\mathbf{H}(\omega)=-\mathbf{N}(\mathbf{M}+i\omega\mathbf{1})^{-1}
\mathbf{N}-\mathbf{1}. \label{eq:Homega}
\end{equation}
Assuming that the outputs and inputs are initially in the vacuum state, this
matrix of frequency responses enables a complete description of the
input-output behavior of the converter.

Now we are interested in the elements of $\mathbf{H}(\omega)$ relating an
input terahertz quadrature to an output optical quadrature, or an input
optical quadrature to an output terahertz quadrature. It turns out that the
non-zero frequency responses relate either input and output $q$ quadratures,
or input and output $p$ quadratures. In fact, the form of the frequency
responses are independent of whether we are considering $q$ or $p$
quadratures. So henceforth we focus exclusively on the $q$ quadratures, with
the understanding that the same results apply to the $p$ quadratures.

Next, the $q$ quadrature terahertz-to-optical frequency response and
optical-to-terahertz frequency response are given by the matrix elements
\begin{equation}
H_{12}(\omega)=\frac{\langle\hat{q}_{\mathrm{o,out}}(\omega)\rangle}%
{\langle\hat{q}_{\mathrm{t,in}}(\omega)\rangle},\ H_{21}(\omega)=\frac
{\langle\hat{q}_{\mathrm{t,out}}(\omega)\rangle}{\langle\hat{q}_{\mathrm{o,in}%
}(\omega)\rangle},
\end{equation}
respectively. It turns out that, aside from the differing reference frames
which have been transformed out of this description, the terahertz-to-optical
frequency response is the same as the optical-to-terahertz frequency response.
Therefore, we may define a single frequency-dependent transmissivity, as
indicated in Eq.~(\ref{eq:tMainText}) of the main text, by
\begin{equation}
t(\omega)\equiv H_{12}(\omega)=H_{21}(\omega),
\end{equation}
Now using Eq.~(\ref{eq:Homega}), we find that
\begin{subequations}
\begin{align}
t(\omega)  &  =\frac{-8g_{\mathrm{o}}g_{\mathrm{t}}\sqrt{\kappa_{\mathrm{o}%
}\kappa_{\mathrm{t}}}}{4g_{\mathrm{t}}^{2}(\kappa_{\mathrm{o}}-2i\omega
)+(\kappa_{\mathrm{t}}-2i\omega)d(\omega)},\\
d(\omega)  &  =4g_{\mathrm{o}}^{2}+(\kappa_{\mathrm{o}}-2i\omega
)(\kappa_{\mathrm{qp}}-2i\omega).
\end{align}
Setting $\omega=0$, corresponding to a terahertz signal at $\omega_{t}$ and an
optical signal at $\omega_{o}$, then leads to the zero-frequency gain quoted
in Eq.~(\ref{eq:tomega0MainText}) of the main text.

In the limit of negligible quasiparticle damping $(\kappa_{qp}\rightarrow0)$,
symmetric couplings $(g_{o}=g_{t}=g)$, and symmetric out-couplings
$(\kappa_{o}=\kappa_{t}=\kappa)$, we have $|t(\omega=0)|\rightarrow1$. The
frequency-dependence of the magnitude and phase response may also be studied
in this limit, in order to ascertain the useful bandwidth of the converter.
They are given by
\end{subequations}
\begin{subequations}
\begin{align}
|t(\omega)|^{2}  &  =\frac{\kappa^{2}}{\kappa^{2}+4\omega^{2}}\frac{16g^{4}%
}{16g^{2}(g^{2}-\omega^{2})+\omega^{2}(\kappa^{2}+4\omega^{2})},\nonumber\\
& \label{eq:magresponse}\\
\phi(\omega)  &  =\tan^{-1}\left[  \frac{\omega(8g^{2}+\kappa^{2}-4\omega
^{2})}{4\kappa(g^{2}-\omega^{2})}\right]  , \label{eq:phiomega}%
\end{align}
respectively. The group delay through the converter, $g(\omega)=d\phi/d\omega
$, is obtained from Eq.~(\ref{eq:phiomega}) as
\end{subequations}
\begin{equation}
g(\omega)=\frac{2\kappa}{\kappa^{2}+4\omega^{2}}+\frac{2\kappa(2g^{2}%
+\omega^{2})}{16g^{2}(g^{2}-\omega^{2})+\omega^{2}(\kappa^{2}+4\omega^{2})}.
\label{eq:groupdelay}%
\end{equation}
To determine the bandwidth across which the magnitude response and group delay
are approximately constant, we Taylor expand Eqs.~(\ref{eq:magresponse}) and
(\ref{eq:groupdelay}) in the (assumed) small parameters $\omega/\kappa$ and
$\omega/g$. This yields the approximations
\begin{subequations}
\begin{align}
|t(\omega)|^{2}  &  =1-\frac{(\kappa^{2}-8g^{2})^{2}}{16g^{4}\kappa^{2}}%
\omega^{2}+\mathcal{O}\left[  (\omega/\kappa)^{4},(\omega/g)^{4}\right]
,\nonumber\\
& \label{eq:TaylorMagnitude}\\
g(\omega)  &  =\frac{8g^{2}+\kappa^{2}}{4g^{2}\kappa}+\left(  \frac{3\kappa
}{8g^{4}}-\frac{8}{\kappa^{3}}-\frac{\kappa^{3}}{64g^{6}}\right)  \omega
^{2}\nonumber\\
&  +\mathcal{O}\left[  (\omega/\kappa)^{3},(\omega/g)^{3}\right]  .
\label{eq:TaylorDelay}%
\end{align}
We would expect that $g\ll\kappa$ in the proposed device, and in this regime
the quadratic terms in Eqs.~(\ref{eq:TaylorMagnitude}) and
(\ref{eq:TaylorDelay}) are negligible provided that
\end{subequations}
\begin{equation}
\omega\ll4g(g/\kappa),
\end{equation}
in addition to the assumed regime, $\omega\ll g,\kappa$. If we assume
$g=10^{8}\,\mathrm{s^{-1}}$ and $\kappa=10^{9}\,\mathrm{s^{-1}}$, then the
bandwidth over which we would expect high-fidelity transmission is of the
order of $\sim10\,\mathrm{MHz}$.



\end{document}